\documentclass[twocolumn,superscriptaddress,longbibliography,
aps,prx,floatfix,preprintnumbers]{revtex4-2}

\usepackage[normalem]{ulem}
\usepackage{graphicx}
\usepackage{bm}
\usepackage{color}
\usepackage{amsmath}
\usepackage{amssymb}
\usepackage{epstopdf}
\usepackage{lipsum}
\usepackage{gensymb}
\usepackage{threeparttable}
\usepackage{multirow}
\usepackage{scrextend}
\usepackage{xfrac}

\usepackage[urlcolor=blue,colorlinks=true,citecolor=blue,linkcolor=blue,pdfstartview={FitH},bookmarks=false]{hyperref}

\newcommand{\emmpp}{Eu{\it M}$_{2}${\it Pn}$_{2}$\;} 

\usepackage[dvipsnames]{xcolor}
\definecolor{mygreen}{rgb}{0.0, 0.6, 0.0}
\definecolor{pjorange}{rgb}{0.8, 0.3, 0.0}
\definecolor{jlblue}{rgb}{0.2, 0.5, 0.7}

\graphicspath{{fig/}{./fig/}{.}}

\begin{document}

\title{Topological character of the antiferromagnetic EuMg$_{2}$Bi$_{2}$}

\author{Mazharul Islam Mondal}
\affiliation{Department of Physics, University of Central Florida, Orlando, Florida 32816, USA}

\author{Issam Mahraj}
\affiliation{Institute of Nuclear Physics, Polish Academy of Sciences, W. E. Radzikowskiego 152, PL-31342 Krak\'{o}w, Poland}

\author{Milo Sprague}
\affiliation{Department of Physics, University of Central Florida, Orlando, Florida 32816, USA}

\author{Sabin Regmi}
\affiliation{Department of Physics, University of Central Florida, Orlando, Florida 32816, USA}
\affiliation{Center for Quantum Actinide Science and Technology, Idaho National Laboratory, Idaho Falls, Idaho 83415, USA}

\author{Xiaxin Ding}
\affiliation{Glenn T. Seaborg Institute, Idaho National Laboratory, Idaho Falls, Idaho 83415, USA}

\author{Firoza Kabir}
\affiliation{Glenn T. Seaborg Institute, Idaho National Laboratory, Idaho Falls, Idaho 83415, USA}

\author{Himanshu Sheokand}
\affiliation{Department of Physics, University of Central Florida, Orlando, Florida 32816, USA}

\author{Krzysztof Gofryk}
\affiliation{Center for Quantum Actinide Science and Technology, Idaho National Laboratory, Idaho Falls, Idaho 83415, USA}

\author{Dariusz Kaczorowski}
\affiliation{Institute of Low Temperature and Structure Research, Polish Academy of Sciences, Ok\'{o}lna 2, 50-422 Wroc\l{}aw, Poland}

\author{Andrzej~Ptok}
\email[e-mail: ]{aptok@mmj.pl}
\affiliation{Institute of Nuclear Physics, Polish Academy of Sciences, W. E. Radzikowskiego 152, PL-31342 Krak\'{o}w, Poland}

\author{Madhab Neupane}
\email[e-mail: ]{Madhab.Neupane@ucf.edu}
\affiliation{Department of Physics, University of Central Florida, Orlando, Florida 32816, USA}

\date{\today}

\begin{abstract}
Antiferromagnetic \emmpp compounds, where {\it M} is a metal element and {\it Pn} is a pnictogen element, have been recognized as candidates for realizing a topologically nontrivial electronic structure.
In this paper, we focus on EuMg$_2$Bi$_2$, whose topological nature still remains unclear.
We present a comprehensive study based on several experimental and theoretical techniques.
Magnetic susceptibility, electrical resistivity, and specific heat capacity measurements confirm the existence of an antiferromagnetic ordering.
The electronic band structure was investigated by high-resolution angle-resolved photoemission spectroscopy (ARPES), supported by {\it ab initio} calculations.
ARPES measurement reveals that the electronic structure of this system is dominated by linearly dispersive hole-like bands near the Fermi level.
Theoretical analyses of the electronic band structure indicates that EuMg$_2$Bi$_2$ is a strong topological insulator, which should be reflected in the presence of a metallic surface state.
We also theoretically examine the magnetic-field-induced anomalous Hall conductivity, confirming previously reported observations.
\end{abstract}

\maketitle

\section{Introduction}

Europium-based \emmpp pnictides, where {\it M} is a metal and {\it Pn} is a pnictogen element, have attracted significant attention due to the possibility of topologically nontrivial electronic structure in the presence of intrinsic magnetic order~\cite{tokura.yasuda.19,bernevig.felser.22,
wang.zhang.23,puthiya.verzola.25,chen.dong.25}.
The magnetic moments provided by the Eu $4f$ states can modify the electronic band structure, giving rise to the realization of new quantum states.
For example, in EuCd$_{2}$As$_{2}$, magnetic moments tuned by external magnetic field can induce nontrivial topological phases such as Dirac or Weyl semimetal~\cite{ma.nie.19,soh.dejuan.19,
wang.jo.19,ma.wang.20,cao.yu.22,taddei.yin.22,sakhya.wang.22,sprague.regmi.24}.
Similarly, the intrinsic magnetic order in EuIn$_{2}$As$_{2}$ or EuSn$_{2}$As$_{2}$ lead to realization of higher-order topological axion insulator state~\cite{xu.song.19,li.gao.19,regmi.hosen.20}.

In this paper we focus on EuMg$_{2}$Bi$_{2}$.
Similar to many \emmpp compounds, it crystallizes in the trigonal CaAl$_{2}$Si$_{2}$-like structure with P$\bar{3}$m1 symmetry (space group No.~164)~\cite{marshall.pletikosic.21}.
A-type antiferromagnetic order~\cite{pakhira.heitmann.21} is observed below the N\'eel temperature $T_{N} \sim 7$~K~\cite{marshall.pletikosic.21}.
In such case, similar to the magnetic topological insulator MnBi$_{2}$Te$_{4}$~\cite{otrokov.klimovskikh.19,gong.guo.19}, ferromagnetic Eu layers are stacked antiferromagnetically along the $c$ axis.
Initial studies of the EuMg$_{2}$Bi$_{2}$ electronic band structure suggested the realization of a Dirac state~\cite{kabir.hosen.19}, while the $\mathbb{Z}_{2}$ invariant indicated magnetic topological features~\cite{marshall.pletikosic.21}.
More recently, a topological Dirac semimetal phase induced by spin--orbit coupling (SOC) has been suggested~\cite{wang.qian.24}.
However, a final consensus regarding the topological character of EuMg$_{2}$Bi$_{2}$ has not yet been reached.

Here, we aim to resolve the puzzle of topological character of EuMg$_{2}$Bi$_{2}$.
We present a comprehensive study of the dynamical and electronic properties of EuMg$_{2}$Bi$_{2}$. 
Thermal and transport measurements confirm a phase transition and antiferromagnetic order with $T_{N} = 6.7$~K.
A theoretical investigation of the lattice dynamics, based on {\it ab initio} methods, shows that the system is dynamically stable, with no imaginary or soft phonon modes.
The electronic band structure was investigated both experimentally, using angle-resolved photoemission spectroscopy (ARPES) and theoretically, using modern {\it ab initio} techniques. 
We identify linearly dispersive bands crossing the Fermi level.
Additionally, theoretical investigation suggests the existence of a gap and features consistent with strong topological insulator.
Nevertheless, similar to previous studies~\cite{marshall.pletikosic.21,kondo.ochi.23,
wang.qian.24}, the experimentally observed Fermi level is shifted by approximately $0.1$~eV relative to theory, which complicates experimental verification of the topological character.
Finally, we examine also magnetic-field-induced anomalous Hall conductivity, confirming previously reported observation.
We link the observed anomalous Hall conductivity to field-induced bands spin-splitting.

The paper is organized as follows.
The methods and techniques used in this study are described in Sec.~\ref{sec.tech}.
In Sec.~\ref{sec.main_prop}, we present the main properties of EuMg$_{2}$Bi$_{2}$,  including the crystal structure (Sec.~\ref{sec.crystal}), lattice dynamics (Sec.~\ref{sec.lattice_dyn}), and magnetic properties (Sec.~\ref{sec.mag_ord}).
Next, in Sec.~\ref{sec.el_properties}, we analyze the electronic properties,  and in Sec.~\ref{sec.topo_prop}, we discuss the topological properties.
Finally, Section~\ref{sec.summary} summarizes our key findings and conclusions.

\section{Methods and techniques}
\label{sec.tech}

\subsection{Experimental details}

{\it Sample growth and characterizations.}---
Single crystals of EuMg$_{2}$Bi$_{2}$ were grown using the Sn flux method as described by in previous work~\cite{canfield.fisk.92}.
The crystal structure was determined using X-ray diffraction with a Kuma-Diffraction KM4 four-circle diffractometer and Mo K$\alpha$ radiation.
The chemical composition was analyzed using energy-dispersive X-ray analysis with an FEI scanning electron microscope equipped with an EDAX Genesis XM4 spectrometer.
The electrical resistivity, heat capacity, and magnetic susceptibility of the crystals were measured using a Quantum Design PPMS system, which included a $9$~T superconducting magnet and utilized the ACT, HC, and VSM measurement options, respectively.

{\it Synchrotron measurements.}---
Synchrotron-based ARPES measurements of the electronic structure of EuMg$_2$Bi$_2$ were performed at the Stanford Synchrotron Radiation Lightsource (SSRL) Endstation 5-2, equipped with a SCIENTA DA30L electron spectrometer.
To ensure a clean surface, the samples were cleaved and maintained under an ultra-high vacuum environment, with a pressure better than $1\times10^{-11}$~Torr at a temperature of around $10$~K. 
ARPES measurements provided an energy resolution better than $20$~meV and an angular resolution finer than $0.2^{\circ}$. 
The stability of EuMg$_{2}$Bi$_{2}$ cleaved surface was maintaned under UHV conditions during the typical $20$-hour measurement period, showing no signs of degradation.

\subsection{Computational details}

First-principles DFT-based calculations were performed using the projector augmented-wave (PAW) potentials~\cite{blochl.94} implemented in the Vienna Ab initio Simulation Package ({\sc Vasp}) code~\cite{kresse.hafner.94,kresse.furthmuller.96,kresse.joubert.99}.
For the exchange-correlation energy, the generalized gradient approximation (GGA) in the Perdew, Burke, and Ernzerhof (PBE) parametrization was used~\cite{perdew.burke.96}.
Similarly to the previous study~\cite{marshall.pletikosic.21}, we introduced the correlation effects on Eu $4f$ orbitals within DFT+U approach, proposed by Dudarev {\it et al.}~\cite{dudarev.botton.98}.
We assume the effective on-site Coulomb interactions $U = 5$~eV and the effective on-site exchange interactions $J = 0.75$~eV.
The energy cutoff for the plane-wave expansion was set to $350$~eV.
Optimization of the structural parameters (in the presence of the SOC and Eu $4f$ electrons treated as a valence states) was performed using $12 \times 12 \times 4$ ${\bm k}$--point grid, using the Monkhorst--Pack scheme~\cite{monkhorst.pack.76}.
As a convergence condition for the optimization loop, we used an energy change below $10^{-6}$~eV and $10^{-8}$~eV for the ionic and electronic degrees of freedom, respectively.

The study of the electronic surface properties was performed by constructing the tight-binding model in the maximally localized Wannier orbitals~\cite{marzari.vanderbilt.97,souza.marzari.01,marzari.mostofi.12}.
The exact band structure was used to construct the tight-binding model by {\sc Wannier90}~\cite{pizzi.vitale.20}.
We constructed two models: one in the absence, and one in the presence of the Eu $f$ states.
For the Eu $4f$ electrons treated as core states, we constructed a $13$-orbital, $26$-band tight-binding model based on Eu $d$, Mg $s$, and Bi $p$ orbitals.
In the absence of the $f$ states, this model does not contain Eu magnetic moments, thus the primitive unit was used.
Simultaneously, for the Eu $4f$ electrons treated as valence states, we constructed a $33$-orbital, $66$-band tight-binding model.
In this case, the Eu magnetic moments associated with $f$ states were included, thus the magnetic unit cell (containing two primitive unit cells) was used.
Thus, the basis, as previously ($13$-orbitals per primitive unit cell), was completed by additional $f$ orbitals. 
For spin $\uparrow$ and $\downarrow$  subspaces, we added seven orbitals centered at Eu with adequate magnetic moments.
Finally, the electronic surface states were calculated using the surface Green's function technique for a semi-infinite system~\cite{lopez.lopez.85}, implemented in {\sc WannierTools}~\cite{wu.zhang.18}.

Dynamic properties were calculated using the direct {\it Parlinski--Li--Kawazoe} method~\cite{parlinski.li.97}, implemented in the {\sc Phonopy} package~\cite{togo.chaput.23,togo.23}. 
Within this method, the interatomic force constants (IFCs) are calculated from the Hellmann--Feynman (HF) forces acting on the atoms after displacements of individual atoms inside the supercell.
We performed these calculations using a supercell corresponding to $3 \times 3 \times 2$ unit cells.
During these calculations, a reduced $3 \times 3 \times 3$ ${\bm k}$-grid was used.

\begin{figure*}
\centering
\includegraphics[width=\linewidth]{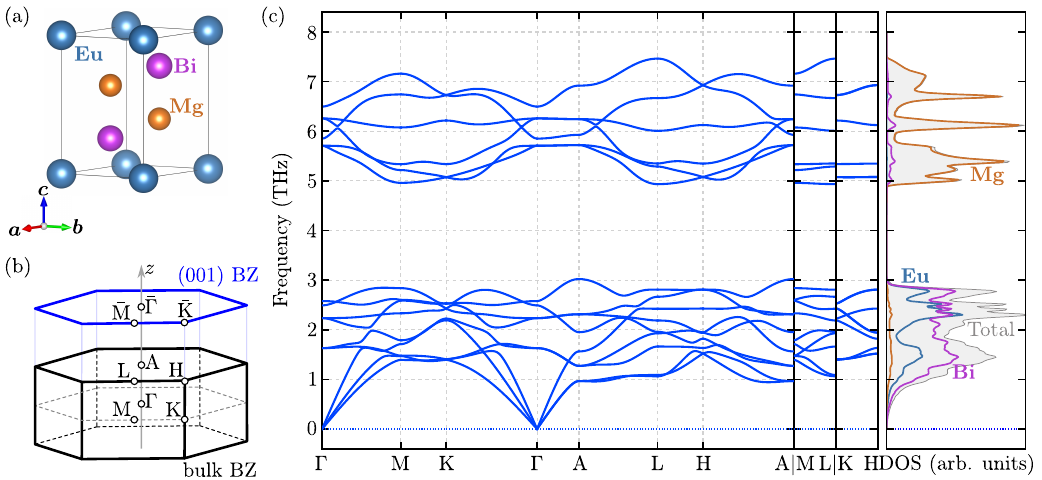}
\caption{
(a) Crystal structure of EuMg$_{2}$Bi$_{2}$ with P$\bar{3}$m1 symmetry, and (b) the corresponding bulk Brillouin zone with its high-symmetry points.
(c) The phonon dispersion curves and the corresponding density of states. 
\label{fig.ph_band}}
\end{figure*}

\begin{table}[!b]
\caption{
Characteristic frequencies (THz) and symmetries of irreducible representations (Irr) of the optical phonon modes at $\Gamma$ point for EuMg$_{2}$Bi$_{2}$.
\label{tab.irr_gamma}
}
\begin{ruledtabular}
\begin{tabular}{ccc}
Frequency & Irr & Activity \\
\hline
1.629 & E$_{g}$ & Raman \\
2.234 & E$_{u}$ & -- \\
2.491 & A$_{1g}$ & Raman \\
2.578 & A$_{2u}$ & IR \\
5.708 & E$_{u}$ & IR \\
5.852 & A$_{1g}$ & Raman \\
6.259 & E$_{g}$ & -- \\
6.497 & A$_{2u}$ & IR \\
\end{tabular}
\end{ruledtabular}
\end{table}

\section{Main properties}
\label{sec.main_prop}

\subsection{Crystal structure}
\label{sec.crystal}

EuMg$_{2}$Bi$_{2}$ crystallizes in the CaAl$_{2}$Si$_{2}$-like structure with trigonal P$\bar{3}$m1 symmetry (space group No.~164).
The unit cell contain five atoms, corresponding to one formula unit [see Fig.~\ref{fig.ph_band}(a)].
The lattice parameters obtained from powder x-ray diffraction are $a = 4.7771$~\AA ~and $c = 7.8524$~\AA, which are in good agreement with previously reported values~\cite{may.mcguire.11,pakhira.tanatar.20,marshall.pletikosic.21}.
From theoretical optimization of the crystal structure, we obtained $a = 4.853$~\AA~and $c = 7.953$~\AA, which are close to the experimental values.
The Eu, Mg, and Bi atoms occupy the Wycoff positions $1a$ ($0$,$0$,$0$), $2d$ ($1/3$,$2/3$,$z_\text{Mg}$), and $2d$ ($1/3$,$2/3$,$z_\text{Bi}$), respectively, where $z_{x}$ are free parameters for Mg and Bi atoms.
The theoretically optimized free parameters are $z_\text{Mg} = 0.6267$ and $z_\text{Bi} = 0.2482$, which are close to experimentally reported~\cite{may.mcguire.11,pakhira.tanatar.20,marshall.pletikosic.21}.

\begin{figure}[!b]
\centering
\includegraphics[width=\linewidth]{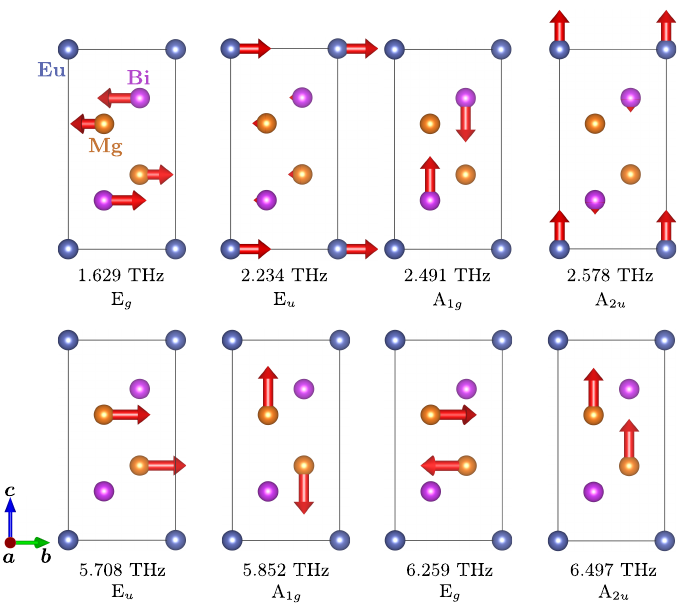}
\caption{
Phonon optical modes at the $\Gamma$ point.
\label{fig.vibr}}
\end{figure}

\subsection{Lattice dynamics}
\label{sec.lattice_dyn}

Phonon dispersion curves and the corresponding density of states are presented in Fig.~\ref{fig.ph_band}(c). 
The system does not exhibit any imaginary (soft) modes, and is therefore dynamically stable.
The acoustic branches exhibit the typical linear dispersion near the $\Gamma$ point.
The upper and lower groups of phonon branches are separated by a large gap of approximately $2$~THz.
Such gap arises from the significant mass difference between heavy (Eu and Bi) and light (Mg) atoms.
For example, such a gap is much smaller or not presented in the phonon spectrum for EuMg$_{2}$P$_{2}$ or EuMg$_{2}$As$_{2}$~\cite{yasin.ullah.25}.

The partial density of states clearly shows a separation between the vibrations of the heavy Eu and Bi atoms and those of light Mg atoms~\cite{shuai.geng.16}.
The high-frequency modes (above $5$~THz) are dominated by Mg atom vibrations, whereas the low-frequency modes (below $3$~THz) arise from mixed Eu and Bi vibrations.
A similar separation has been reported for EuZn$_{2}$P$_{2}$~\cite{rybicki.komendera.24}.
However, in contrast to EuMg$_{2}$Bi$_{2}$, the vibrations of the light P pnictogen atom in EuZn$_{2}$P$_{2}$ occupy the high-frequency range (above $6.5$~THz), while heavier Zn  metal atom contributes primarily to low-frequency range (below $4.8$~THz).
This clearly shows the important role of the constituent elements in shaping the phonon spectra.

The phonon modes at the $\Gamma$ point can be decomposed into the following irreducible representations:
\begin{eqnarray}
\Gamma_{acoustic} &=& \text{A}_{2u} + \text{E}_{u} , \\
\nonumber \Gamma_{optic} &=& 2 \text{A}_{1g} + 2 \text{A}_{2u} + 2 \text{E}_{u} + 2 \text{E}_{g} .
\end{eqnarray}
Here, the A$_{1g}$ and E$_{g}$ modes are Raman active, while the A$_{2u}$ and $E_{u}$ are infra-red active.
the characteristic frequencies are collected in Tab.~\ref{tab.irr_gamma}, and schematic representations of the vibrations are shown in Fig.~\ref{fig.vibr}.
As expected, the two-dimensional E-modes corresponds to vibrations within the $ab$ plane, whereas, the A-modes involve out-of-plane vibrations along the $c \parallel z$ direction.

\begin{figure}[!t]
\centering
\includegraphics[width=\linewidth]{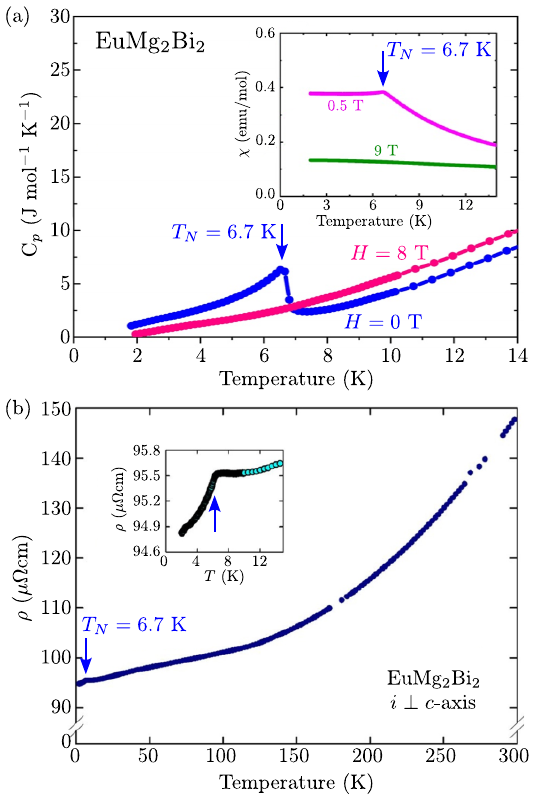}
\caption{
(a) Heat capacity of EuMg$_{2}$Bi$_{2}$ measured in zero magnetic field (blue curve) and in a high field of $9$~T (red curve). 
A pronounced anomaly marks a phase transition at $6.7$~K. 
The inset shows the temperature dependence of the magnetic susceptibility of EuMg$_{2}$Bi$_{2}$, with the antiferromagnetic transition at $T_{N} = 6.7$ K. 
Magnetic susceptibility measurements were performed under $0.5$~T (violet curve) and $9$~T (green curve) manetic fields. 
(b) Temperature variation of the electrical resistivity of EuMg$_{2}$Bi$_{2}$, measured within the trigonal plane. 
\label{fig.el_trans}
}
\end{figure}

\begin{figure}[!t]
\centering
\includegraphics[width=\linewidth]{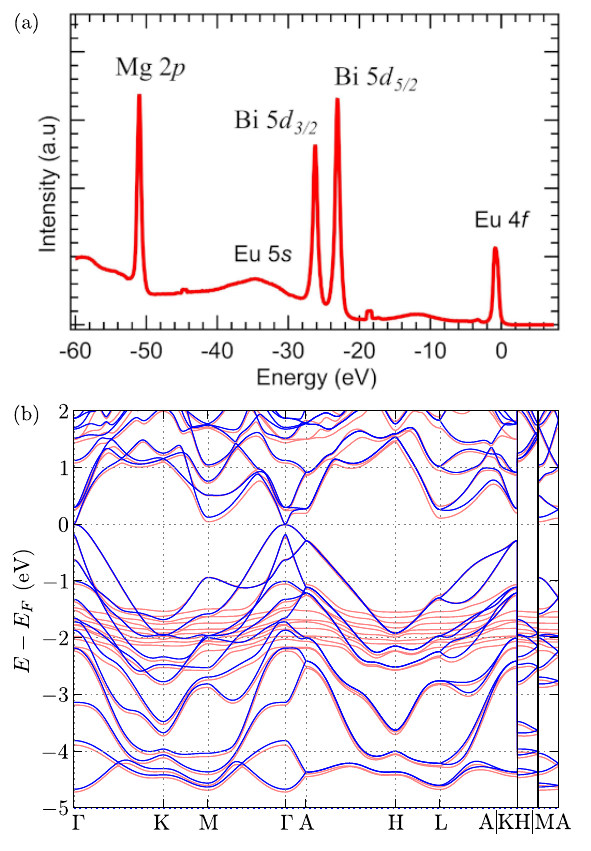}
\caption{
(a) Measured core levels spectrum of EuMg$_{2}$Bi$_{2}$.
Sharp peaks correspond to the Eu $5s$, Eu $4f$, SOC split Bi $5d$, and Mg $2p$ levels (as labeled).
(b) Calculated electronic band structure along high-symmetry directions.
Results are shown with spin--orbit coupling included, for Eu $4f$ electrons treated as core states (blue lines) and valence states (pink lines).
\label{fig.el_band}
}
\end{figure}

\begin{figure*}
\centering
\includegraphics[width=\linewidth]{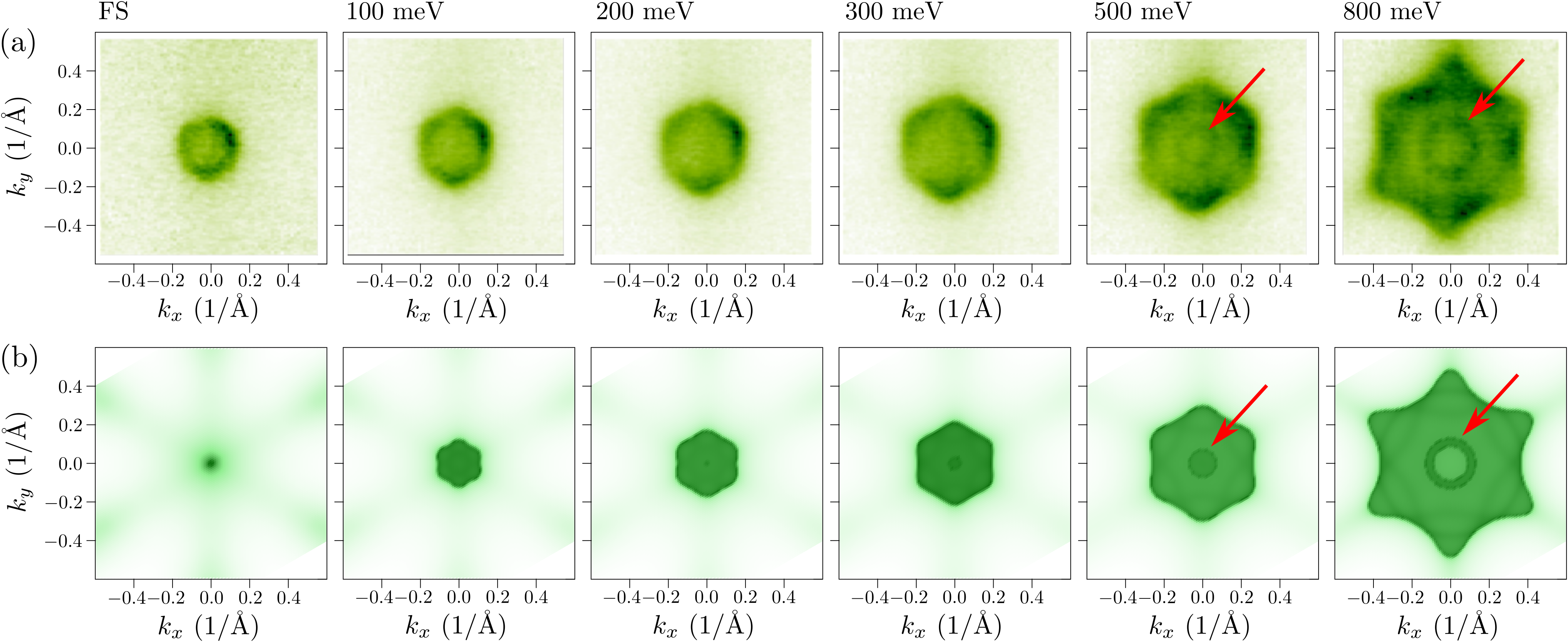}
\caption{
Fermi surface map and constant-energy contours of EuMg$_{2}$Bi$_{2}$.
(a) Experimentally measured Fermi surface maps acquired at photon energy of $60$~eV and constant-energy contour plots at various binding energies
The corresponding binding energies are noted in the plots. 
All measurements were performed at the SSRL beamline 5-2 at a temperature of $10$~K in the Paramagnetic (PM) phase.
(b) Corresponding theoretically obtained results (bottom panels). 
\label{fig.cecp}
}
\end{figure*}

\subsection{Magnetic order}
\label{sec.mag_ord}

EuMg$_{2}$Bi$_{2}$ is known to be an A-type antiferromagnet, in which spins are aligned ferromagnetically within interplane, while adjacent layers are aligned antiferromagnetically~\cite{pakhira.heitmann.21}.
In colinear calculations without SOC, the antiferromagnetic phase has a lower energy than the ferromagnetic phase, while an energy difference of approximately $1.44$~meV/f.u.

Transition to the AF state at $T_{N} = 6.7$~K is well visible in  the heat capacity, magnetic susceptibility and longitudinal resistivity measurements (Fig.~\ref{fig.el_trans}).
The heat capacity measurements reveal a distinct $\lambda$-like anomaly at $T_{N}$, indicating the occurence of a second-order phase transition~\cite{may.mcguire.11}.
The presence of long-range AF ordering is further corroborated by magnetic susceptibility measurements. 
As shown in the inset of Fig.~\ref{fig.el_trans}(a), a pronounced anomaly is observed at the phase transition, consistent with previous reports~\cite{may.mcguire.14, may.mcguire.11}.
Application of a strong magnetic field suppresses this feature.
The electrical resistivity measurements presented in Fig.~\ref{fig.el_trans}(b) show that $\rho (T)$ exhibits metallic behavior, with electrical resistivity increasing as the temperature rises. 
Consistent with the heat capacity and magnetic susceptibility results, the electrical resistivity also displays a characteristic anomaly at the antiferromagnetic phase transition.

Below $T_{N}$, the Eu magnetic moments lie in the $ab$ plane.
Such magnetic moment configuration was confirmed by the neutron diffraction measurements~\cite{pakhira.heitmann.21}.
Nevertheless, we should notice, that the predicted electronic properties can strongly depend on the  direction of the magnetic moments (as will be discussed later in Sec.~\ref{sec.el_properties}).
The Eu$^{2+}$ ions ($S = 7/2$, $L = 0$) have an average magnetic moment of $5.3$~$\mu_{B}$ at $T = 4$~K, which is smaller than the expected $7$~$\mu_{B}$ because the moments are not yet fully saturated to its full values at $T = 0$~K~\cite{pakhira.heitmann.21}.
In fact, theoretical calculation give value closer to the nominal $6.95$~$\mu_{B}$.
Indeed, saturated value in the presence of magnetic field is around $7$~$\mu_{B}$ at $T = 2$~K~\cite{pakhira.tanatar.20,marshall.pletikosic.21}.

\section{Electronic properties}
\label{sec.el_properties}

The core-level photoemission spectrum is presented in Figure~\ref{fig.el_band}(a).
The Mg $2p$, Bi $5d$, and Eu $4f$ states are well visible as sharp peak, whereas the Eu $5p$ states are more broadly distributed.
Due to strong SOC, the Bi $5d$ states are split into Bi $5d_{3/2}$ and Bi $5d_{5/2}$ states.

\begin{figure*}
\centering
\includegraphics[width=\linewidth]{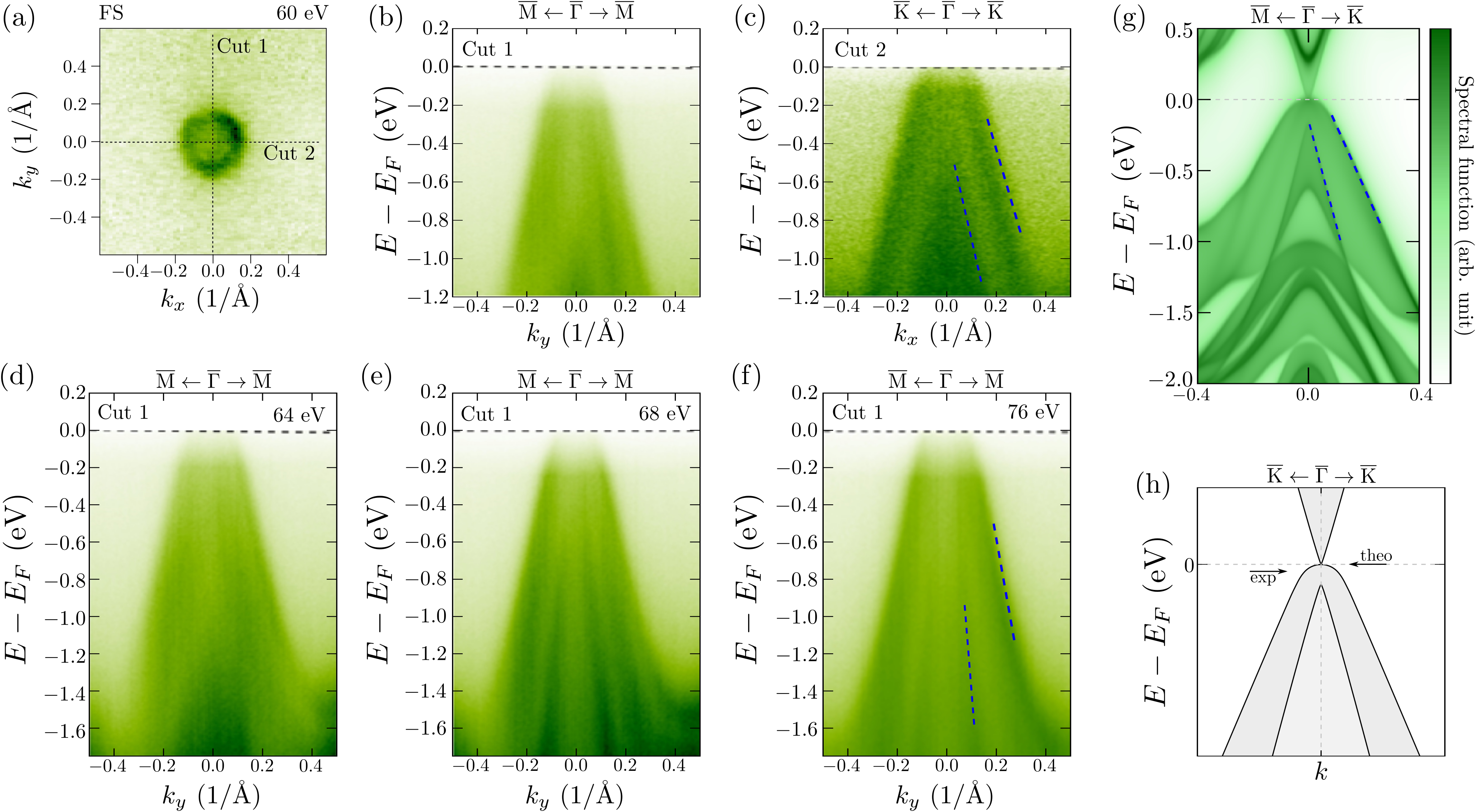}
\caption{
Observation of electronic structure along various high-symmetry directions in momentum space. 
(a) Fermi surface taken at a photon energy of 60 eV. 
The black dashed lines represent different cut directions along which the dispersion maps were taken. 
Energy dispersion along (b) cut 1 (direction $\overline{\text{M}}$--$\overline{\Gamma}$--$\overline{\text{M}}$, and (c) cut 2 (direction $\overline{\text{K}}$--$\overline{\Gamma}$--$\overline{\text{K}}$). 
(d-f) Photon-energy-dependent measurements along $\overline{\text{M}}$--$\overline{\Gamma}$--$\overline{\text{M}}$ direction (cut 1).
Blue dashed lines indicate linearly dispersive hole bands.
(g) Theoretically obtained spectral function along the $\overline{\text{M}}$--$\overline{\Gamma}$--$\overline{\text{K}}$ path.
(h) Schematic representation of the bands at the $\overline{\Gamma}$ points along $\overline{\text{K}}$--$\overline{\Gamma}$--$\overline{\text{K}}$. 
The arrows indicate the theoretical and experimental Fermi energies.
\label{fig.fermi}
}
\end{figure*}

A comparison of the bulk electronic band structure for Eu $4f$ states treated as core and valence states is presented in Fig.~\ref{fig.el_band}(b).
The Eu $4f$ states are well visible at a binding energy of approximately $1.5$~eV as nearly-dispersionless flat bands. 
It should be noted that results obtained using an effective Coulomb interactions of $U - J = 4.25$~eV reproduce the Eu $4f$ states well [cf.~Fig.~\ref{fig.el_band}(a)].
Similar to other Eu-based compounds, such as EuZn$_{2}$P$_{2}$~\cite{krebber.kopp.23}, EuAl$_{2}$Ge$_{2}$~\cite{pakhira.kundu.23}, EuCd$_{2}$Sb$_{2}$~\cite{su.gong.20} or
EuCd$_{2}$As$_{2}$~\cite{ma.nie.19,ma.wang.20,nelson.king.24}, the Eu $4f$ states are localized around a binding energy of $1.75$~eV.
Introducing the Eu $4f$ states does not affect the shape of the electronic bands structure near the Fermi level.
Around the $\Gamma$ point, an electron-like conduction band and a hole-like valence band are well visible.

\subsection{Angle-resolved photoemission spectroscopy}

Comparison with the calculated band structure was carried out by performing ARPES measurements.
First, the Fermi surface and several constant-energy contours (CECs), ranging from $0$ to $800$~meV below the Fermi level, are shown in Fig.~\ref{fig.cecp}.
The ARPES measurements were performed on the naturally cleaving (001) surface at a temperature of $10$~K, which lies in the paramagnetic phase; therefore, the Eu $4f$ states should not play an important role due to the absence of magnetic ordered phase.
Ultraviolet incident photon energy of $60$~eV  were used for the experimental CECs presented here.
The Fermi surface consists of a small circular pocket [Fig.~\ref{fig.cecp}(a)].
This pocket grows with increasing binding energy, indicative of hole-like nature of the associated bands. 
At higher binding energies, this pocket evolves into a six-fold symmetric, bent hexagonal-shaped pocket, accompanied by a circular pocket touches the bent part of the hexagonal-shaped pocket.
At binding energy above approximately $500$~meV, another hole-like circular pocket emerges around the $\overline{\Gamma}$ point [marked by red arrows].

Next, we analyzed the electronic band dispersion cuts along high-symmetry directions (Fig.~\ref{fig.fermi}).
Cuts were taken along the $\overline{\text{M}}$--$\overline{\Gamma}$--$\overline{\text{M}}$ [Fig.~\ref{fig.fermi}(b)] and $\overline{\text{K}}$--$\overline{\Gamma}$--$\overline{\text{K}}$ [Fig.~\ref{fig.fermi}(c)] directions, corresponding to cut 1 and cut 2 along the black dashed lines in Fig.~\ref{fig.fermi}(a).
These cuts reveal two nearly-linearly dispersing hole-like bands along both high-symmetry directions [marked by the blue dashed line {in Fig.~\ref{fig.fermi}(c)].

Moreover, the lower band displays hole-like curvature in its dispersion below $0.2$~eV binding energy. 
The two outer bands are most clearly resolvable along the both $\overline{\text{M}}$--$\overline{\Gamma}$--$\overline{\text{M}}$ and $\overline{\text{K}}$--$\overline{\Gamma}$--$\overline{\text{K}}$ directions where the outer band forms the circular Fermi surface, which intercepts the Fermi level at a radius of about $0.1$~\AA$^{-1}$. 
The inner pocket disperses roughly in parallel with the outer pocket, but appears to reach its maximum just below the Fermi level [marked by red arrow on Fig.~\ref{fig.cecp}].

In Figs.~\ref{fig.fermi}(d)-(f), we present photon-energy-dependent ARPES measurements along the $\overline{\text{M}}$--$\overline{\Gamma}$--$\overline{\text{M}}$ direction.
Notably, the outer pocket exhibits consistent and distinct visibility across all photon energies, indicating weak k$_z$ dispersion, particularly at lower binding energies. 
On the other hand, the inner band shows more pronounced dispersion along k$_{z}$.
Nevertheless, the nearly-linear dispersion of both bands remains clearly visible over a wide photon-energy range.
The experimentally observed band dispersion shows excellent agreement with the DFT-based calculations [cf.~Fig.~\ref{fig.fermi}(g)].

\begin{figure}[!b]
\centering
\includegraphics[width=\linewidth]{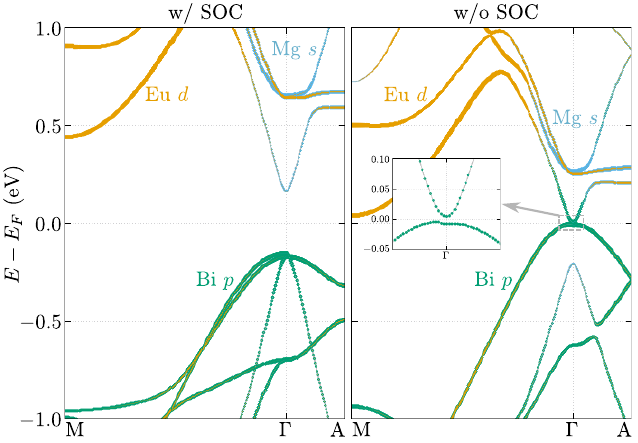}
\caption{
Orbital contributions to the electronic band structure in the absence and presence of the spin--orbit coupling (as labeled).
\label{fig.el_orb}
}
\end{figure}

\subsection{Spin--orbit coupling and Fermi level shift}

We note that the electronic band structure of EuMg$_{2}$Bi$_{2}$ strongly depends on the SOC~\cite{choudhury.mohanta.23}.
This sensitivity arises from the relatively strong SOC of Eu ($140$~meV and $30$~meV for $p$ and $f$ orbitals, respectively) and Bi ($165$~meV for $p$ states).
The role of the SOC becomes evident when examining the orbital projections of the bands near the Fermi level (Fig.~\ref{fig.el_orb}).
The valence (conduction) band is mostly composed of Bi $p$ states (Eu $d$ and Mg $s$ states).
Thus, the inclusion of SOC leads to strong band decoupling [see also Fig.~\ref{fig.el_soc} in the Supplemental Material (SM)~\footnote{See Supplemental Material at [URL will be inserted by publisher] for additional theoretical and experimental results.}], which aslo affect the observed gap.
A previous study suggested that the gap closes~\cite{wang.qian.24}.
However, precise calculations show that the gap does not close [see inset in Fig.~\ref{fig.el_orb}(b)]; instead, it decreases from $316$~meV to $13$~meV.

Unfortunately, the experiential results clearly show the existence of electronic states at the Fermi level.
This feature is visible in the ARPES measurements, where a Fermi surface is observed [see Fig.~\ref{fig.cecp}]~\cite{marshall.pletikosic.21,kondo.ochi.23,
wang.qian.24}.
The existence of a conduction channel is further supported by the low-temperature resistivity measurements presented in Sec.~\ref{sec.mag_ord}.
However, the theoretical results discussed earlier show that ``ideal'' EuMg$_{2}$Bi$_{2}$ should behave as an insulator with the Fermi level located inside the band gap, [Fig.~\ref{fig.fermi}(h)].
A similar discrepancy has previously been reported for EuCd$_{2}$As$_{2}$ and EuZn$_{2}$As$_{2}$, where theoretical calculations place the Fermi level inside the band gap, whereas APRES measurements show it located near the top of the valence band~\cite{ma.nie.19,ma.wang.20,jo.kuthanazhi.20, 
cao.yu.22}.
In conclusion, comparison of the experimental and theoretical band structure indicates that the experimental Fermi level is shifted by approximately $0.1$~eV relative to the theoretical value.

\begin{figure}[!b]
\centering
\includegraphics[width=\linewidth]{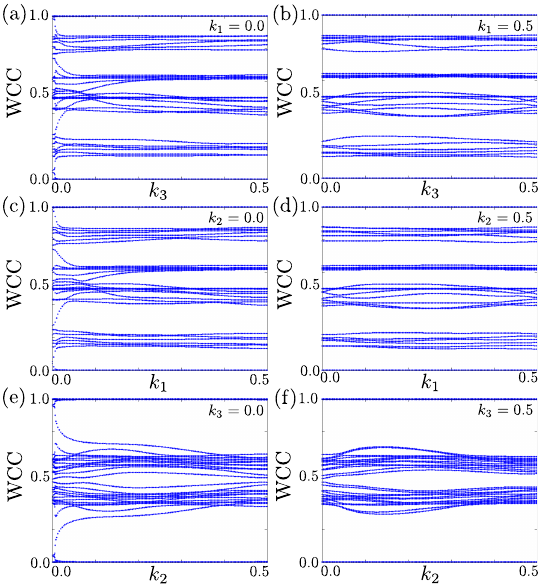}
\caption{
Evolution of the Wannier charge center (WCC) for different planes in the Brillouin zone.
\label{fig.z2}
}
\end{figure}

\section{Topological properties}
\label{sec.topo_prop}

In order to identify the possible topological nature of EuMg$_{2}$Bi$_{2}$, we calculated the $\mathbb{Z}_{2}$ invariant by employing the Wannier charge center (WCC) presented in Fig.~\ref{fig.z2}.
Here $\mathbb{Z}_{2} = ( \nu_{0} ; \nu_{1} \mu_{2} \nu_{3} )$, where $\nu_{i} \equiv \nu ( k_{i} = \pi )$ are weak topological indexes, while $\nu_{0}$ is the strong topological index, given as mod~$2$ of sum of all $\nu ( k_{i} = 0 )$ and $\nu ( k_{i} = \pi )$ indexes~\cite{fu.kane.07b}.
In our case, we observe a significant difference in WCC evolution for $k_{i} = 0$ and $k_{i} = \pi$ planes (cf.~left and right panels in Fig.~\ref{fig.z2}).
In fact, from our calculations we found $\nu ( k_{i} = 0 ) = 1$, and $\nu ( k_{i} = \pi ) = 0$, which correspond to $\nu_{0} = 1$ and $\mathbb{Z} = ( 1 ; 0 0 0 )$.
These results are in agreement with initial study presented in Ref.~\cite{marshall.pletikosic.21}.
Moreover, similar predictions have been reported for the isostructural SmMg$_{2}$Bi$_{2}$~\cite{kundu.pakhira.22}.

Therefore, EuMg$_{2}$Bi$_{2}$ should be classified as a strong topological insulator.
Such topological properties can be associated with SOC-induced band inversion at the $\Gamma$ point, visible in Fig.~\ref{fig.el_orb}.
In the absence of SOC, the strong contribution of Bi $p$ orbitals is visible only in the valence band.
However, in the presence of the strong SOC, the Bi $p$ contribution is also possible in the conduction bands.
For the strong topological insulator, metallic surface states closing the gap can be expected independently of the realized surface.
Indeed, theoretical calculations indicate possible surface states (Fig.~\ref{fig.el_ss}).
Similar results were obtained for the isostructural SmMg$_{2}$Bi$_{2}$, confirming that possibility~\cite{kundu.pakhira.22}.
Unfortunately, due to the previously discussed shift of the Fermi level, confirmation of this prediction is impossible by our ARPES measurements. 
Nevertheless, the Fermi level can be tuned by, e.g., potassium dosing.
Such a technique was successfully used in the case of EuCd$_{2}$As$_{2}$ to confirm the absence of the Weyl points~\cite{nelson.king.24}.

\begin{figure}[!b]
\centering
\includegraphics[width=\linewidth]{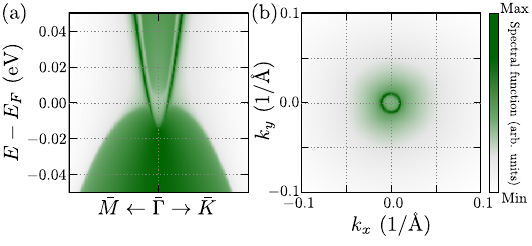}
\caption{
(a) The electronic surface state around $\bar{\Gamma}$ point and (b) corresponding constant energy contour on the theoretical Fermi level.
\label{fig.el_ss}
}
\end{figure}

\subsection{External magnetic field inducing topological phase}

The external magnetic field can also affect the topological properties.
For example, the magnetic field (${\bm H}$) can induce the anomalous Hall (AH) effect.
Such observation was reported by Kondo {\it et al.} in Ref.~\cite{kondo.ochi.23}.
For ${\bm H} \parallel c$, the long-range magnetically ordered phase exists up to $4$~T~\cite{pakhira.tanatar.20}.
Below this field, the AH conductivity increases from zero to a saturated value $\rho_{xy}^{A} \approx 50$~$\Omega^{-1}$cm$^{-1}$.
The authors claim that the observation of non-zero AH conductivity for magnetic field along $c$ direction is possible due to the band splitting induced by the Eu magnetic moments.
In such case, the initially in-plane magnetic moment start, in the presence of an external magnetic field, to realize a canted antiferromagnetic states, with a non-zero Eu net magnetization.
Above the saturation field, all magnetic moments are parallel to $c$ and form the ferromagnetic phase.
Non-vanishing total magnetization gives rise to non-zero AH conductivity.
A similar mechanism was suggested for EuCd$_{2}$As$_{2}$~\cite{cao.yu.22}, EuIn$_{2}$As$_{2}$~\cite{yan.jiang.22}, and EuZn$_{2}$As$_{2}$~\cite{yi.zheng.23,regmi.blawat.25}.

\begin{figure}[!t]
\centering
\includegraphics[width=\linewidth]{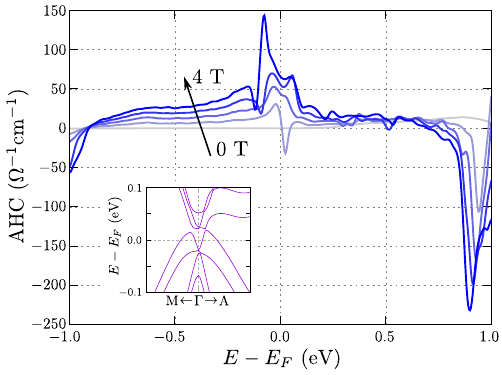}
\caption{
Theoretical magnetic-field dependence of anomalous Hall conductivity (AHC).
Inset presents electronic band structure around the $\Gamma$ point in the presence of magnetic field of $4$~T.
\label{fig.el_ahc}
}
\end{figure}

In fact, the external magnetic field can affect the system in two ways.
First, by the aforementioned modification of the magnetic moment directions.
However, this ``local'' effect should be associated only with the Eu magnetic moments (i.e., $4f$ states).
Second, the ordinary modification of all bands by the Zeeman-like splitting. 
This is a global effect, and affects all bands independently of the orbital character.
In this spirit, we perform the AH conductivity calculations a function of the external magnetic field.
To mimic the Zeeman effect, the external magnetic field was introduced in the calculations as an additional spin-dependent on-site term in the Hamiltonian.
The AH conductivity is given by standard formula, as a Brillouin zone integral of the Berry curvature over the occupied bands~\cite{yao.kleinman.04,wang.yates.06,nagaosa.sinova.10}:
\begin{eqnarray}
\sigma_{xy}^{AHC} = - \frac{e^{2}}{\hbar} \int_{BZ} \frac{d{\bm k}}{(2\pi)^{2}} \Omega_{z} \left( {\bm k} \right) ,
\end{eqnarray}
where $\Omega_{z} \left( {\bm k} \right)$ is total Berry curvature.

Results of our calculations are presented in Fig.~\ref{fig.el_ahc}.
As we can see, the non-zero AH conductivity emerges with the external magnetic field increasing.
In fact, the non-zero Berry curvature is guaranteed by the band spin-splitting induced by the external magnetic field.
The complex band mixing and several avoided band crossings (see inset in Fig.~\ref{fig.el_ahc}) give rise to a significant increase of the AH conductivity~\cite{wang.yates.06}.
Surprisingly, the largest positive AH conductivity is expected around energies $-0.1$~eV, as reported in Ref.~\cite{pakhira.tanatar.20}.

\section{Summary}
\label{sec.summary}

In summary, in this paper we discuss the electronic properties of the EuMg$_{2}$Bi$_{2}$ antiferromagnet (T$_N$ = 6.7 K) in the context of its topological features.
To study the electronic band structure, we performed systematic high-resolution ARPES measurements and modern {\it ab initio} techniques. 
Our measurements reveal the presence of hole-like pockets at the $\Gamma$ point, constructed by linearly dispersing bands.
Such observations were reproduced by the theoretical study.
Constant energy contours show the existence of states crossing the Fermi level, which is in contradiction with theoretical results showing the topological nature of the EuMg$_{2}$Bi$_{2}$.
However, this problem was reported previously and is related to the shift of the experimental Fermi level with respect to theoretical one, obtained for ``ideal'' crystal structure.

Theoretical investigation indicate that EuMg$_{2}$Bi$_{2}$ a strong topological insulator with topological index $\mathbb{Z}_{2} = ( 1; 0 0 0 )$.
In such case, metallic surface states should emerge independently of the realized surface.
Unfortunately, due to previously discussed shift of the Fermi level, confirmation of this prediction is impossible by our ARPES measurements. 
Nevertheless, the Fermi level can be tuned by, e.g., potassium dosing, which was used  in the case of EuCd$_{2}$As$_{2}$ to confirm the absence of the Weyl points~\cite{nelson.king.24}.
We believe that similar technique can be used in the future investigations.

Finally, we investigate the magnetic-field-induced anomalous Hall effect. 
Previous study suggest a predominant role of the rare-earth atom magnetic moments.
We show that the observed non-zero anomalous Hall conductivity in the presence of an external magnetic field, can also be associated with magnetic-field-induced band spin-splitting.

We believe the intrinsic magnetism and possible topological features make EuMg$_{2}$Bi$_{2}$ an attractive platform for further study of interplay between magnetism and nontrivial topology.

\begin{acknowledgments}
Some figures in this work were rendered using {\sc Vesta}~\cite{momma.izumi.11} software.
M.N. is supported by the DOE Office of Science, Basic Energy Sciences (BES), under Award No. DE-SC0024304. 
F.K. and X.D. acknowledge support from Idaho National Laboratory's Laboratory Directed Research and Development (LDRD) program under DOE Idaho Operations Office Contract DE-AC07-05ID14517. 
K.G. and S. R. acknowledge support from the Division of Materials Science and Engineering, Office of Basic Energy Sciences, Office of Science of U.S. Department of Energy.
A.P. is grateful to Laboratoire de Physique des Solides in Orsay (CNRS, University Paris Saclay) for hospitality during a part of the
work on this project.
This work was supported by National Science Centre (NCN, Poland) under Projects No.~2021/43/B/ST3/02166 (A.P.) and No.~2021/41/B/ST3/01141 (D.K.).
\end{acknowledgments}

\bibliography{biblio.bib}


\clearpage
\newpage

\onecolumngrid

\begin{center}
\textbf{\Large Supplemental Material}\\[.3cm]
\textbf{\large Topological character of the antiferromagnetic EuMg$_{2}$Bi$_{2}$}\\[.3cm]
Mazharul Islam Mondal$^{1}$, Issam Mahraj$^{2}$, Milo Sprague$^{1}$, Sabin Regmi$^{1,3}$, Xiaxin Ding$^{4}$, Firoza Kabir$^{4}$, Himanshu Sheokand$^{1}$, Krzysztof Gofryk$^{3}$, Dariusz Kaczorowski$^{5}$, Andrzej~Ptok$^{2}$, and Madhab Neupane$^{1}$ \\[.2cm]
{\itshape
${}^{1}$Department of Physics, University of Central Florida, Orlando, Florida 32816, USA \\
${}^{2}$Institute of Nuclear Physics, Polish Academy of Sciences, W. E. Radzikowskiego 152, PL-31342 Krak\'{o}w, Poland \\
${}^{3}$Center for Quantum Actinide Science and Technology, Idaho National Laboratory, Idaho Falls, Idaho 83415, USA \\
${}^{4}$Glenn T. Seaborg Institute, Idaho National Laboratory, Idaho Falls, Idaho 83415, USA \\
${}^{5}$Institute of Low Temperature and Structure Research, Polish Academy of Sciences, Ok\'{o}lna 2, 50-422 Wroc\l{}aw, Poland \\
}
(Dated: \today)
\\[0.3cm]
\end{center}

\setcounter{equation}{0}
\renewcommand{\theequation}{S\arabic{equation}}
\setcounter{figure}{0}
\renewcommand{\thefigure}{S\arabic{figure}}
\setcounter{section}{0}
\renewcommand{\thesection}{S\arabic{section}}
\setcounter{table}{0}
\renewcommand{\thetable}{S\arabic{table}}
\setcounter{page}{1}


In this Supplemental Material, we present additional results:
\begin{itemize}
\item Sec.~\ref{sec.s0} -- Additional theoretical results.
\item Sec.~\ref{sec.s1} -- Electronic structure of (100) surface.
\item Sec.~\ref{sec.s2} -- Polarization dependent APRES measurement.
\end{itemize}

\newpage

\section{Additional theoretical results}
\label{sec.s0}

\begin{figure}[!hb]
\centering
\includegraphics[width=0.75\linewidth]{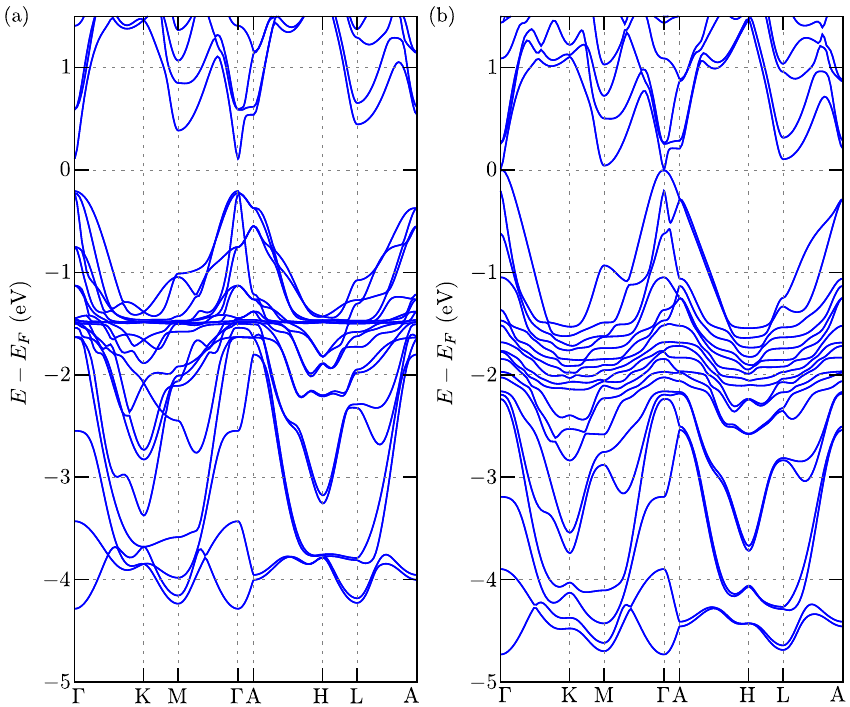}
\caption{
The electronic band structure in the absence (a) and presence (b) of the spin--orbit coupling.
\label{fig.el_soc}
}
\end{figure}

\begin{figure}[!hb]
\centering
\includegraphics[width=0.6\linewidth]{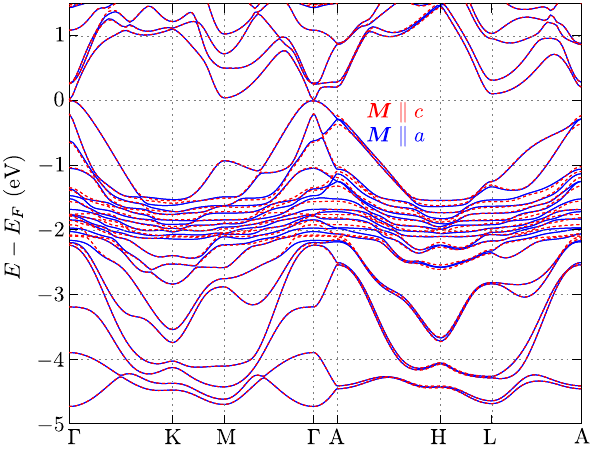}
\caption{
The electronic band structure obtained for Eu magnetic moments along $a$ and $c$ axis (solid blue and dashed red line, respectively).
Result in the presence of the spin--orbit coupling.
\label{fig.el_afm}
}
\end{figure}

\newpage

\section{Electronic Structure Along Alternate (100) Cleave}
\label{sec.s1}

The main text presents the electronic band dispersion along the natural (001) cleaving direction where linearly dispersive hole-like bands are observed near the Fermi level.
To complement these results, we also present the corresponding dispersion along the alternate (100) cleaving plane, revealing the full three-dimensional dispersion in EuMg$_2$Bi$_2$. 
Fig.~\ref{fig.s3} reveals this dispersion, as measured using both $84$~eV and $98$~eV photons, both with a linear horizontal (LH) polarization. 
The top four panels of Fig.~\ref{fig.s3} demonstrates the localized hole-like nature of the Fermi pockets surrounding the $\overline{\Gamma}$-point. 
We have overlaid the (100) first Brillouin zone (BZ) on the Fermi surface presented in the top-right panel. 
The BZ was calculated directly from the experimental lattice constants and is clearly commensurate with the observed periodicity along the $\overline{\text{K}}$--$\overline{\Gamma}$--$\overline{\text{K}}$ direction.
The bottom left two panels [Fig.~\ref{fig.s3}(e,f)] show the dispersion measured along the $\overline{\text{K}}$--$\overline{\Gamma}$--$\overline{\text{K}}$ and $\overline{\text{A}}$--$\overline{\Gamma}$--$\overline{\text{A}}$ high-symmetry directions using an incident photon energy of $84$~eV, where we clearly resolve linearly dispersive bands. 
Comparing Fig.~\ref{fig.s3}(e,f) we observe a striking similarity in the dispersion along these two directions, despite the layered atomic arrangement in this compound. 
Fig.~\ref{fig.s3}(g,h) present the same dispersion cuts using a higher photon energy. 
At $98$ eV, the matrix element effects allow for increased visibility of the neighboring BZs. 
Here we see that along $\overline{\text{A}}$--$\overline{\Gamma}$--$\overline{\text{A}}$ Fig.~\ref{fig.s3}(g) the hybridization of electrons across neighboring BZs appears to be relatively restricted, leading to the high isotropy of the pockets within $\sim -400$ meV of the Fermi energy.

\begin{figure}[!ht]
\centering
\includegraphics[width=\linewidth]{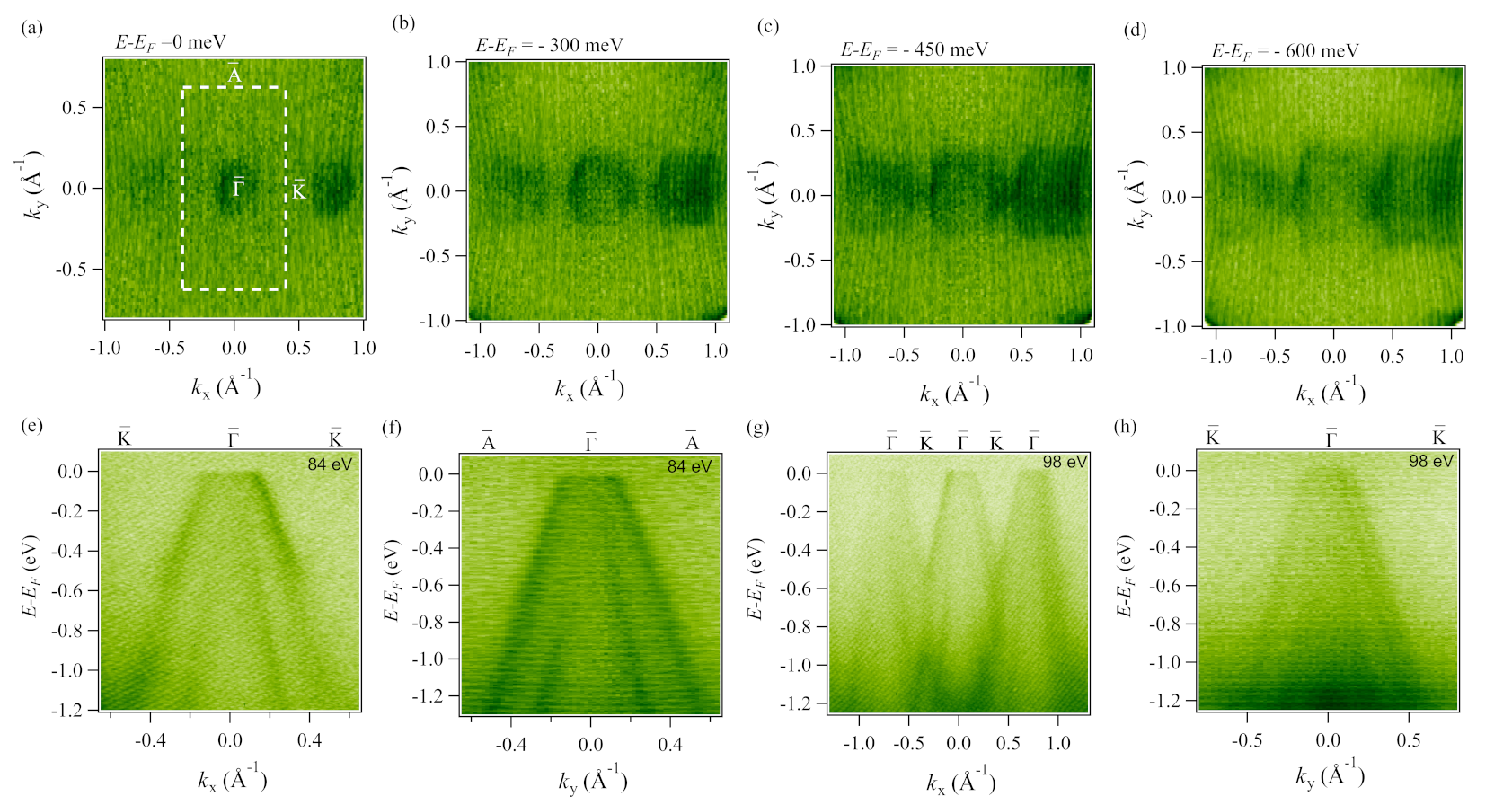}
\caption{Electronic band dispersion along the $\overline{\text{K}}$--$\overline{\Gamma}$--$\overline{\text{K}}$ and $\overline{\text{A}}$--$\overline{\Gamma}$-$-\overline{\text{A}}$ directions. (a-d) Fermi surface and constant energy contours taken at binding energies of $0$~meV, $-300$~meV, $-450$~meV, and $-600$~meV, respectively. The surface-projected Brillouin zone along the (100) direction is overlaid upon the Fermi surface in panel (a), indicating the orientations of the $\overline{\text{A}}$ and $\overline{\text{K}}$ high symmetry points. (e,f) Electronic band dispersion along the $\overline{\text{K}}$--$\overline{\Gamma}$--$\overline{\text{K}}$ and $\overline{\text{A}}$--$\overline{\Gamma}$--$\overline{\text{A}}$ directions measured using $84$~eV incident photons with a linear horizontal (LH) polarization. (g,h) The same high-symmetry cuts with higher photon energy, $98$~eV.
\label{fig.s3}}
\end{figure}

\newpage

\section{Polarization-dependence of ARPES-measured dispersions}
\label{sec.s2}

The Bi $p$ orbital composition of the valence band structure has been invoked in the interpretation of the remarkable thermoelectric properties in EuMg$_2$Bi$_2$. 
To corroborate this assignment, we have varied the polarization of incident photons in our ARPES study. 
Fig.~\ref{fig.s4} presents the variations of the measured dispersion spectra upon changing linear polarization, which allows for the identification of symmetric/antisymmetric behavior of the orbitals across the photoemission mirror plane. 
Fig.~\ref{fig.s4}(a,b) shows both the LH and linear vertical (LV) spectra, taken along the $\overline{\text{A}}$--$\overline{\Gamma}$--$\overline{\text{A}}$ direction, using $84$~eV photon energy. 
We observe the suppression of the ARPES intensity with LV-polarized photons, which has the vector potential lie perpendicular to the photoemission plane. 
This indicates strongly antisymmetric orbital composition about the photoemission plane, corroborating the $p$-orbital character of the valence states.

\begin{figure}[!ht]
\centering
\includegraphics[width=0.75\linewidth]{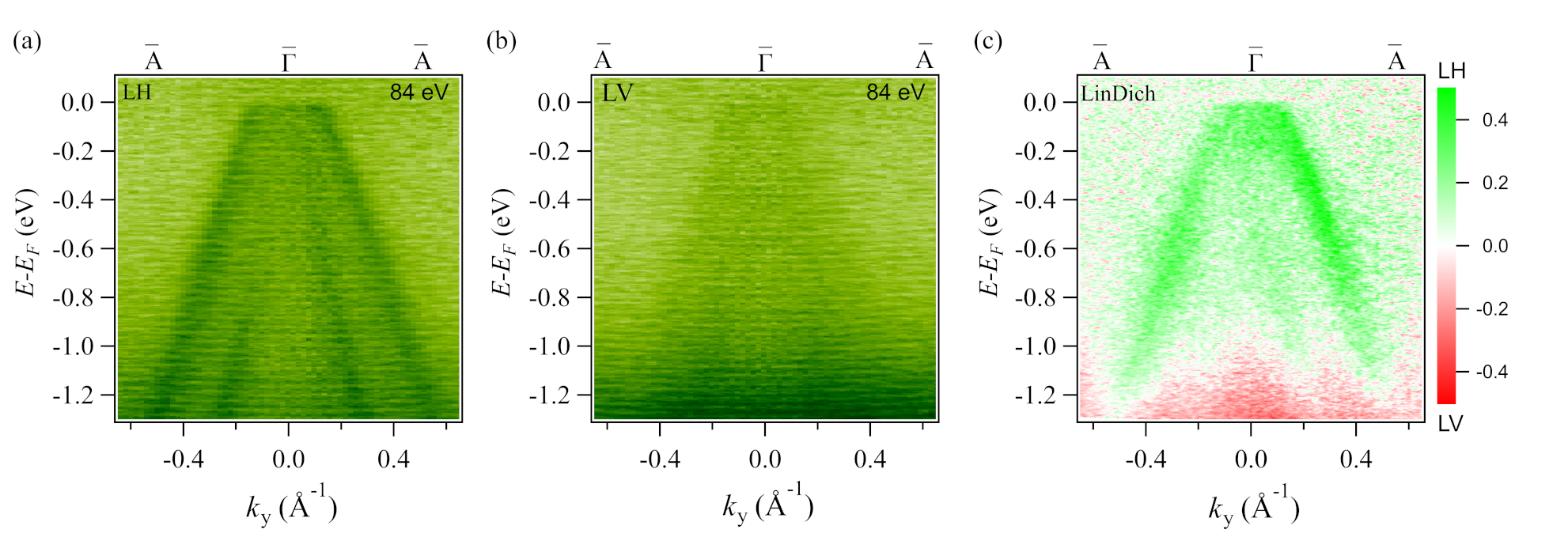}
\caption{Dependence on incident photon linear polarization. The $\overline{\text{A}}$--$\overline{\Gamma}$--$\overline{\text{A}}$ direction is measured using linear horizontal (LH) and linear vertical (LV) incident photons are shown in the left and middle panels, respectively. The right panel presents the normalized difference between the LH and LV-measured spectra.
\label{fig.s4}}
\end{figure}

Additionally, we have repeated these measurements using right- and left-handed circularly polarized incident photons, as indicated by CR/CL in Fig.~\ref{fig.s5}.
We observe a relatively modest change in the photoemission intensities between the top and bottom rows along both the $\overline{\text{A}}$--$\overline{\Gamma}$--$\overline{\text{A}}$ and $\overline{\Gamma}$--$\overline{\text{K}}$--$\overline{\Gamma}$--$\overline{\text{K}}$--$\overline{\Gamma}$ directions. 
This indicates a non-zero orbital angular momentum \cite{park.kim.12}.

\begin{figure}[!ht]
\centering
\includegraphics[width=\linewidth]{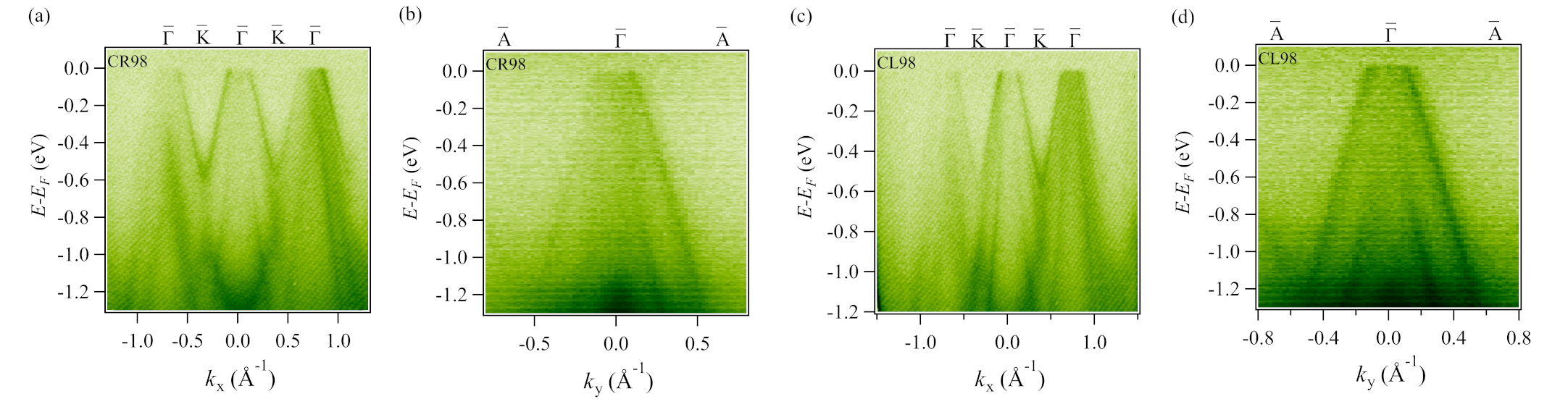}
\caption{Comparison of right (top row) and left (bottom row) handed circularly polarized incident photons. The $\overline{\text{K}}$--$\overline{\Gamma}$--$\overline{\text{K}}$ dispersion is shown on the left panels and the $\overline{\text{A}}$--$\overline{\Gamma}$--$\overline{\text{A}}$ dispersion is shown on the right. A photon energy of $98$ eV was used for all presented data.
\label{fig.s5}}
\end{figure}

\end{document}